\documentclass{aa}  

\usepackage{graphicx}
\usepackage{hyperref}
\usepackage[varg]{txfonts}
\usepackage{caption}
\usepackage{subcaption} 
\usepackage{multicol}
\usepackage{booktabs}

\usepackage{placeins}

\usepackage{mathtools} 
\DeclarePairedDelimiterXPP\BigOSI[2]%
  {\mathcal{O}}{(}{)}{}%
  {\SI{#1}{#2}}

\usepackage{float}
\usepackage{amsmath}
\usepackage{xcolor}

\usepackage{pifont} 

\usepackage{lipsum} 

\hypersetup{
  colorlinks=true,   
  urlcolor=blue,     
  linkcolor=blue,     
  citecolor=blue
}

\begin{document}

\title{Signatures of Fuzzy Dark Matter Inside Radial Critical Curves}

\author{J. M. Palencia    \thanks{palencia@ifca.unican.es}
          \inst{1}       
          \and
          Paloma Morilla\inst{2,3}
          \and
          Sung Kei Li\inst{4}
          \and
          J. M. Diego\inst{1}
          \and
          Amruth Alfred\inst{4}
          \and
          Thomas J. Broadhurst\inst{2,3,5}
          \and
          B. J. Kavanagh\inst{1}
          \and 
          Jeremy Lim\inst{4}
          }
   \institute{Instituto de Física de Cantabria (CSIC-UC), Avda. Los Castros s/n, 39005 Santander, Spain
      \and
   Department of Physics, University of the Basque Country (UPV/EHU), 48080 Bilbao, Spain
   \and
   Donostia International Physics Center, DIPC, Basque Country, San Sebasti\'an, 20018, Spain
   \and
   Department of Physics, The University of Hong Kong, Pokfulam Road, Hong Kong
   \and
    Ikerbasque, Basque Foundation for Science, Bilbao, Spain
     }


\abstract{We investigate the strong gravitational lensing properties of fuzzy dark matter (FDM) halos,  focusing on the magnification properties near radial critical curves (CCs). Using simulated lenses we compute magnification maps for a range of axion masses and halo configurations. We show that FDM produces enhanced central magnification and secondary CCs that are not easily reproduced by standard cold dark matter (CDM), even when including subhalos. The strength and scale of these effects depend primarily on the de Broglie wavelength, governed by the axion and halo masses. We find that axion masses in the range $m_\psi \sim 10^{-22}$–$10^{-21}\,\mathrm{eV}$ in galaxy-mass halos lead to distinctive magnification distributions. Our results suggest that observations of highly magnified, compact sources near radial arcs, such as quasars or supernovae, could serve as a powerful test for the presence of FDM.
}

\keywords{Gravitational lensing, Gravitational lensing: strong, Dark matter, Galaxies, Galaxy: halo}

\maketitle
\section{Introduction} \label{sec:intro}
The standard $\Lambda$CDM cosmological model provides the most successful description of our Universe to date~\citep{Planck_cosmo2020}. In this framework, the energy density of the Universe consists predominantly of mysterious dark energy ($\sim$70\%), which drives the accelerated expansion (observed primarily at $z < 1$). The remaining $\sim$30\% corresponds to matter, of which only about 5\% is the familiar baryonic matter, while the remaining $\sim$25\% is attributed to the elusive cold dark matter (CDM). Although the gravitational effects of CDM have been evident for decades, its true nature remains a mystery in modern cosmology. While the $\Lambda$CDM model accurately reproduces the large-scale structure of the Universe, it still faces significant challenges on small scales~\citep{DelPopolo2017}, particularly at scales $\lesssim 1$ Mpc.
These small-scale problems include, among others, the core–cusp problem~\citep{Moore1994, Moore1999, Flores1994, Oh2011, Gentile2004}, the missing satellites problem~\citep{Moore1999, Klypin1999}, and the too-big-to-fail problem~\citep{Boylan2011, Boylan2012}. 

Some progress has been made in addressing the core–cusp tension through high-resolution cosmological simulations of disk galaxies that incorporate strong baryonic feedback, particularly supernova, driven outflows and high star formation thresholds~\citep{Guedes2011, McCarthy2012, Brook2012}. However, the extent to which these mechanisms can alleviate other small-scale issues, such as core formation or satellite abundances, remains debated~\citep{Schaller2015}.

The fundamental nature of dark matter remains unknown. Alternative scenarios to CDM, such as self-interacting dark matter~\citep{Tulin2018} or warm dark matter~\citep{Viel2013}, have been proposed to account for persistent discrepancies between the predictions of standard CDM and observations on small scales. Modifications to the $\Lambda$CDM paradigm have also been proposed to address these small-scale issues. Among the leading CDM candidates are weakly interacting massive particles (WIMPs) \citep{Arcadi2018}, which are fermionic and behave as discrete particles, and wave dark matter ($\psi$DM), a bosonic alternative composed of ultra-light particles with masses below 10~eV~\citep{Hui:2021tkt, Ferreira2021}. This latter model reflects the particle–wave duality of quantum mechanics, as the de Broglie wavelength of the particles exceeds their average separation in galaxies, allowing the dark matter to be treated effectively as a classical wave.

In particular, fuzzy dark matter (FDM) is a kind of $\psi$DM model that is non-self-interacting, non-relativistic, and extremely light, with a mass in the range of $10^{-23}$–$10^{-20}$~eV~\citep{HuBarkana, Hui:2016ltb}. One of the key advantages of FDM is its ability to address two of the aforementioned problems: the core–cusp problem, through the formation of a solitonic core in galaxies~\citep{Schive2014, Mocz2017}, and the missing satellite problem, via the suppression of structure formation below the de Broglie scale~\citep{Robles:2014ysa, Schive:2015kza, Kulkarni:2020pnb}. This suppression is reminiscent of warm dark matter models. However, FDM is still CDM and is indistinguishable from the standard CDM predictions on large scales~\citep{Schive2014, Hui:2021tkt}. 

The FDM prediction of solitonic cores and the suppression of small-scale structure in the linear regime~\citep{Hui:2021tkt} make this model highly testable and subject to numerous current constraints. Observations of the early universe~\citep{Zhang2024}, axion detection experiments~\citep{Aja2022,Leung2019}, and searches for astrophysical-related phenomena~\citep{Eberhardt2024,Pinetti2025} have been used to place general limits on the existence of $\psi$DM. In the case of FDM, the usual mass range, around $10^{-22}\,\mathrm{eV}$, has been constrained by the small-scale spatial fluctuations measured via the Lyman-$\alpha$ forest~\citep{Irsic2017,Kobayashi2017,Armengaud2017}, by galaxy dynamics and internal structure~\citep{Nadler2021} and by the sizes and stellar radial velocities of some ultra-faint dwarf galaxies~\citep{Dalal2022}. These constraints disfavour an axion mass near the commonly adopted value of $10^{-22}$ eV, yet they are subject to systematic uncertainties. These arise from limited data and poorly understood baryonic physics  ~\citep{Hui:2021tkt} or from the lack of full wave simulations that could capture higher-order effects~\citep{Dalal2022}. Properly accounting for these factors could potentially relax the current bounds. Regardless of the strength of these constraints, it is crucial to develop complementary approaches based on independent methods to better assess and mitigate systematic uncertainties.

Gravitational lensing presents itself as an excellent tool to probe axion masses and test the viability of FDM as a dark matter candidate~\citep{Laroche:2022pjm, Amruth2023,Powell2023,Broadhurst2025}. Since lensing traces the underlying projected mass distribution, it is sensitive to substructure on different scales, as predicted by various dark matter models. FDM with axion masses of around $10^{-22}$ eV are particularly notable for producing de Broglie wavelengths that give rise to mass density fluctuations on pc to kpc scales in galaxy clusters and galaxy-scale lens systems, respectively. These fluctuations lead to distinctive patterns in the mass distribution compared to the conventional smooth global profiles expected from standard CDM. Such differences have already been proposed as solutions for discrepancies between observational data and the best-fitting canonical CDM-based lensing models~\citep{Amruth2023}, like the long-standing flux-ratio anomalies~\citep{Keeton2003,Goldberg2010,Xu2015,Shajib2019}, the position anomalies in radio observations~\citep{Spingola2018,Hartley2019}, or the asymmetry in microlensed stars in galaxy clusters~\citep{Broadhurst2025}.

In this paper, we develop a lensing framework tailored to FDM distributions in galaxy-scale systems, with a particular focus on deviations in their magnification patterns compared with the case of standard CDM halos. We focus on radial critical curves (CCs), which are often neglected in the literature, as the smooth models predict demagnified central images, and radial images are also less common than their tangential counterparts. However, certain small-scale objects (smaller than the de Broglie wavelength), such as Active Galactic Nuclei, Quasi-Stellar Objects (QSOs), or Supernovae, can be bright enough to be observed even at modest magnification factors and be sensitive to the magnification changes due to FDM fluctuations. But perhaps the most important fact is that near the CCs, the differences between the standard CDM and FDM can be accentuated. The radial critical region corresponds to the portion of the lens where $(1-\kappa)+\gamma=\varepsilon \approx 0$, where $\kappa$ is the convergence, $\gamma$ is the shear, and $\varepsilon$ is an arbitrarily small number. In this region, since $\gamma>0$ always, then it must be satisfied that $\kappa>1$. In the classical CDM scenario, adding substructure will increase the value of $\kappa$, making the term $(1-\kappa)$ even more negative, and in most cases resulting in a reduction of the magnification ($\mu \propto \varepsilon^{-1}$). In contrast, in FDM, we also have negative mass fluctuations (with respect to the mean), relaxing the condition for criticality and increasing the probability of larger magnification (smaller $|\varepsilon|$). In other words, this simple reasoning leads us to expect a higher number of highly magnified objects in the vicinity of the radial CC region in FDM models compared to standard CDM. In this work, we provide a quantitative assessment of the differences between canonical CDM and FDM near radial CCs in terms of their different magnification statistics. 
Here, we demonstrate that FDM predicts a magnification distribution near the centre of halos, specifically around radial CCs, that differs significantly from that predicted by standard CDM, even when including subhalos as small scale perturbers. This behaviour represents an effect unique to FDM, providing a promising avenue to test FDM models as an alternative to canonical CDM.

This paper is structured as follows. Sect.~\ref{sec:lensing} introduces the basics of gravitational lensing, with a particular emphasis on magnification, the observable we propose to use for distinguishing a FDM Universe from one governed by the classical representation of CDM. In Sect.~\ref{sec:mass profiles}, we describe the various mass profiles implemented in our simulations, which are later compared in terms of their magnification properties. In Sect.~\ref{sec:methodology}, we present the methodology followed in this work. The results of our analysis, highlighting the differences among models and the tests employed, are presented in Sect.~\ref{sec:results}. In Sect.~\ref{sec:discussion}, we discuss the implications of our findings and the potential of magnification as a tool for discriminating between dark matter models. Finally, our main conclusions are summarised in Sect.~\ref{sec:conclusion}. 
We assume the Planck 18 cosmological model~\citep{Planck_cosmo2020} with $\Omega_m=0.31$, $\Lambda=0.69$, and $h=0.676$ (100 km s$^{-1}$ Mpc$^{-1}$). In this work, we study the differences between FDM and the standard CDM description within the $\Lambda$CDM framework. Without loss of generality, we will refer to the latter simply as CDM throughout the paper, even though FDM is itself a form of CDM.

\section{Lensing formalism} \label{sec:lensing}
In this section, we briefly introduce the gravitational lensing formalism~\citep{Schneider1992}, focusing on the gravitational lensing effect caused by galaxy lenses. These lenses are typically described by a Navarro–Frenk–White (NFW) dark matter halo~\citep{Navarro1996,Navarro1997}, consistent with both observations and hydrodynamic simulations (in CDM). A S\'ersic profile is adopted for the contribution from baryons. Finally, for the FDM mass profiles, we include the soliton structure and density fluctuations on top of an NFW profile. 

The formalism presented here is nonetheless mass-model independent and can be readily applied to any deflecting structure, including but not limited to NFW or S\'ersic profiles. In addition, when working with complex mass models, we can take advantage of the deflection angle linearity, i.e., the total deflection angle can be expressed as the sum of the deflection angle of each component. In other words, if the total mass distribution can be expressed as 
\begin{equation}\label{eq:sigmas_lin}
    \Sigma_{\rm tot}({\boldsymbol{\theta}})=\sum_{i=1}^N\Sigma_{i}({\boldsymbol{\theta}}),
\end{equation}
then the total deflection angle is 
\begin{equation}\label{eq:alphas_lin}
    \alpha_{\rm tot}({\boldsymbol{\theta}})=\sum_{i=1}^N\alpha_{i}({\boldsymbol{\theta}}),
\end{equation}
where $\alpha_{i}({\boldsymbol{\theta}})$ is the deflection angle at the position ${\boldsymbol{\theta}}$ generated by $\Sigma_{i}$.

The position of a lensed image ${\boldsymbol{\beta}}$ and the corresponding source position ${\boldsymbol{\theta}}$ are connected through the lens equation, \begin{equation} \label{eq:lens_eq} \boldsymbol{\beta} = \boldsymbol{\theta} - \boldsymbol{\alpha}\left(\Sigma,\boldsymbol{\theta}\right), \end{equation} where $\boldsymbol{\alpha}$ is the deflection angle induced at $\boldsymbol{\theta}$ by a lens with surface mass density $\Sigma(\boldsymbol{\theta})$. Since $\boldsymbol{\alpha}$ depends on $\boldsymbol{\theta}$, the equation is generally non-linear, often admitting multiple image positions for a given source location and lacking an analytical solution in most cases.

The deflection angle is derived from the effective lensing potential:
\begin{equation} \label{eq:eff_potential} \psi(\boldsymbol{\theta}) = \frac{D_{\rm ds}}{D_{\rm d}D_{\rm s}}\frac{2}{c^2}\int\phi(D_{\rm d}\boldsymbol{\theta}, z)\,\rm{d} z,
\end{equation}
where $D_{\rm d}$, $D_{\rm s}$ and $D_{\rm ds}$ represent the angular diameter distances to the lens, to the source, and between the lens and source, respectively. The function $\phi$ denotes the Newtonian potential of the lens. The deflection is then given by
\begin{equation}
\boldsymbol{\alpha} = \boldsymbol{\nabla}_{\boldsymbol{\theta}} \psi,
\end{equation}
and the Laplacian of $\psi$ is related to the surface mass density through
\begin{equation} \label{eq:phi_conv}
\boldsymbol{\nabla}_{\boldsymbol{\theta}}^2 \psi = 2\frac{D_{\rm d}D_{\rm ds}}{D_{\rm s}}\frac{4\pi G}{c^2}\Sigma(\boldsymbol{\theta}) = 2\frac{\Sigma(\boldsymbol{\theta})}{\Sigma_{\rm crit}} \equiv 2\kappa(\boldsymbol{\theta}),
\end{equation}
with $\kappa$ being the convergence, defined as the dimensionless surface mass density relative to the critical value $\Sigma_{\rm crit}$.

Gravitational lensing also modifies the shape and size of background sources. These distortions are encoded in the Jacobian matrix
\begin{equation} \label{eq:jacobian1}
\tens A \equiv \frac{\partial\boldsymbol{\beta}}{\partial\boldsymbol{\theta}} = \delta_{ij} - \frac{\partial \alpha_i}{\partial \theta_j} = \delta_{ij} - \frac{\partial^2 \psi}{\partial \theta_i \partial \theta_j} = \delta_{ij} - \psi_{ij} = \tens M^{-1}, \end{equation}
which is the inverse of the magnification tensor $\tens M$. 

\section{Mass profiles} \label{sec:mass profiles}
In the previous section we introduced the lensing formalism by a generic mass distribution $\Sigma(\boldsymbol{\theta})$, which is the 2D projection of a generic $\rho(\boldsymbol{\theta}, z)$ 3D profile, obtained after integrating the density along the line of sight $z$. In this section, we will summarize the mass distributions used in this work for both the particle description of CDM and the wave-like FDM models.

\subsection{Navarro-Frenk-White profile}
Simulations have shown that, under the $\Lambda$CDM paradigm, the power spectrum of the density perturbations, combined with the collisionless nature of CDM, point to cuspy dark halos with density profiles scaling as $\rho_{\rm DM}(r) \propto r^{-1}$. These profiles are well described by the NFW parametrization~\citep{Navarro1996,Navarro1997}, from very massive galaxy clusters ($M\sim10^{15}M_\odot$) to galaxies with masses around $M\sim10^{11}$–$10^{12}M_\odot$. Simulations of FDM halos also show an NFW profile modulated by a central solitonic core and density fluctuations characteristic of its wave nature \citep{Schive2014}. 

The NFW profile is a radial function that goes as
\begin{equation} \label{eq:NFW}
\rho_{\rm NFW}(r)=\frac{\rho_s}{(r/r_s)(1+r/r_s)^2},
\end{equation}
where the two parameters, $\rho_s$ and $r_s$, are the characteristic density and the scale radius of the halo. Lower-mass halos are usually well described by NFW profiles too, but the core–cusp problem arises: the central density tends to an almost constant value rather than rising steeply at the innermost radii. FDM offers a natural solution to this discrepancy through solitonic structures, as seen in numerical simulations~\citep{Schive2014,Liao2024}.

The characteristic density $\rho_s$ can also be expressed as a function of the concentration parameter:
\begin{equation} \label{eq:rho_s}
\rho_s=\frac{200}{3}\rho_{\rm crit}\frac{c^3}{\left[\ln{(1+c)} - \frac{c}{1+c} \right]}.
\end{equation}

The concentration parameter $c$ is simply the ratio between the radius $r_{200}$ and the scale radius of the halo ($c = r_{200}/r_s$), and it scales with the mass of the halo roughly as $M^{-0.1}$~\citep{Dutton2014}. $r_{200}$ is the radius that encloses an average density 200 times that of the critical density:
\begin{equation} \label{eq:r200}
r_{200} = \frac{1.63\times10^{-2}}{(1+ z)h}\left(\frac{M_{200}}{h^{-1}M_\odot}\right)^{1/3} \left[\frac{\Omega_0}{\Omega(z)}\right]^{-1/3}~{\rm kpc}.
\end{equation}

$M_{200}$ is then the mass enclosed within the sphere of radius $r_{200}$, also commonly used to parametrise the halo, analogously to $\rho_s$ and $r_s$.

Defining $x \equiv r/r_s$, the simplicity of the NFW profile, specifically its radial symmetry, allows for an analytical derivation of its surface mass density:
\begin{equation} \label{eq:NFW_surface_mass_density}
    \Sigma_{\rm NFW}(x) = \frac{2\rho_s r_s}{x^2 - 1}
    \left\{
    \begin{array}{ll}
        1 - \frac{2}{\sqrt{x^2 - 1}} \arctan\sqrt{\frac{x - 1}{x + 1}}, & x > 1 \\
        1 - \frac{2}{\sqrt{1 - x^2}} \operatorname{arctanh}\sqrt{\frac{1 - x}{1 + x}}, & x < 1, \\
    \end{array}
    \right.
\end{equation}
The limit at $x = 1$ is finite: $\Sigma_{\rm NFW}(1) = \frac{2}{3} \rho_s r_s$.
The deflection angle is then given by
\begin{equation} \label{eq:NFW_alpha}
    \alpha_{\rm NFW}(x) = \frac{4\rho_s r_s \Sigma_{\rm crit}^{-1}}{x} \, h(x)
\end{equation}
with the auxiliary function $h(x)$ defined as

\begin{equation} \label{eq:NFW_h}
    h(x) =
        \ln\left(\frac{x}{2}\right) + 
        \begin{cases}
            \frac{2}{\sqrt{x^2 - 1}} \arctan\sqrt{\frac{x - 1}{x + 1}}, & x > 1 \\
            \frac{2}{\sqrt{1 - x^2}} \operatorname{arctanh}\sqrt{\frac{1 - x}{1 + x}}, & x < 1 \\
            1, & x = 1
        \end{cases}
\end{equation}

The mass profile presented here is valid for CDM halos and subhalos according to CDM predictions. FDM also uses it as the central baseline model; the further differences will be introduced in the following subsection.

\subsection{FDM profile}
Simulations show that FDM halo profiles are composed of three distinct components. First, a baseline NFW-like profile, as described in the previous subsection. Second, a central solitonic core that produces a flat inner density profile and naturally addresses the core–cusp problem. Together, these two components constitute the smooth mass distribution of the halo. Third, small-scale wavelike density fluctuations, arising from the quantum nature of the field, introduce granularity beyond the smooth profile. While component (i) was discussed previously, we now focus on components (ii) and (iii), which are unique features of FDM.

\subsubsection{Soliton}
\begin{figure}[tpb]
  \includegraphics[width=\columnwidth]{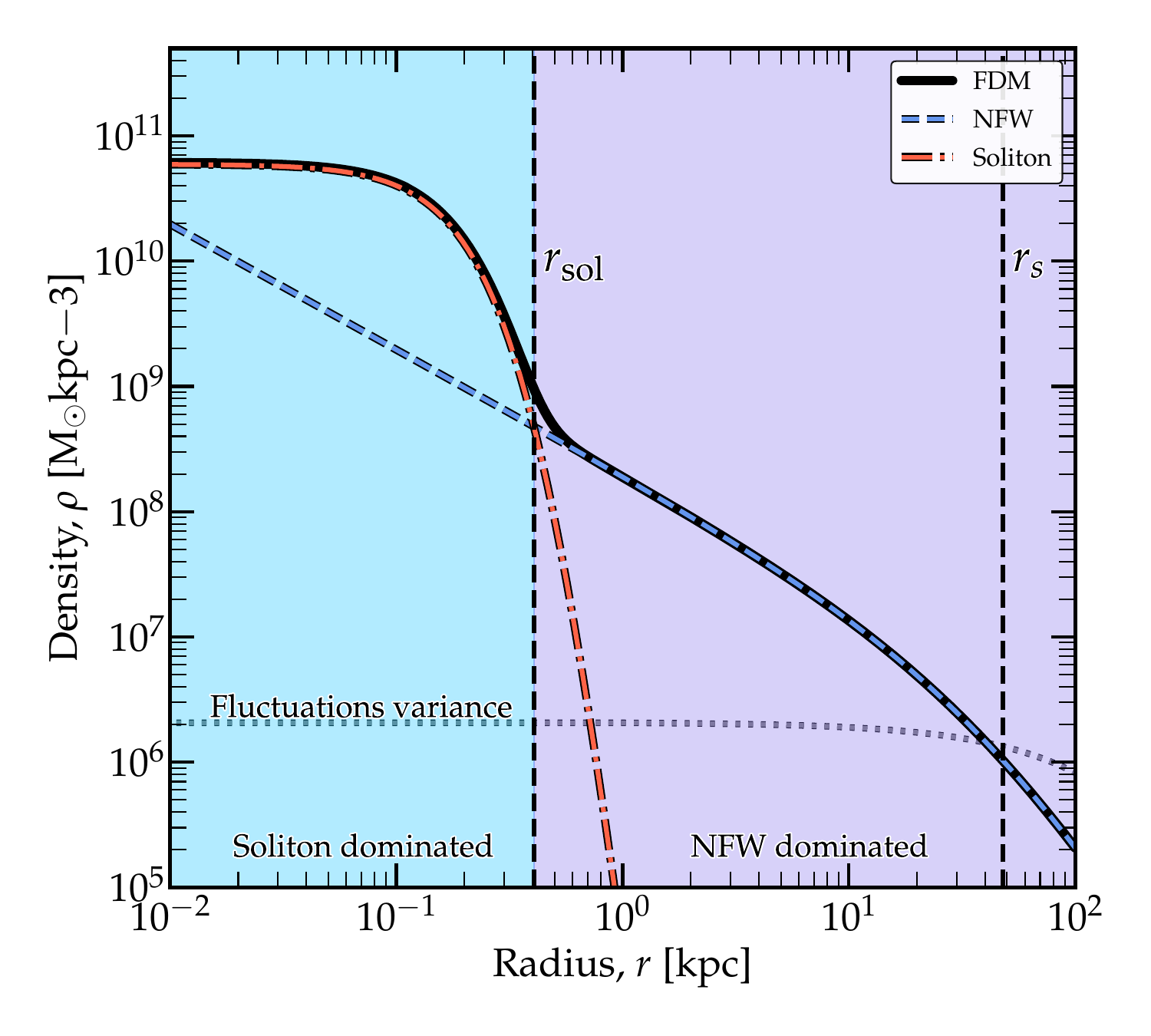} 
      \caption{Radial profile of a FDM halo ($M_h=7\times10^{11}M_\odot$ and $m_\psi=10^{-22}\mathrm{eV}$ at $z=0.9$) neglecting quantum fluctuations (solid black line). The inner profile of the halo is dominated by the soliton structure (red dot-dashed line). At $r = r_{\rm sol}$, it transitions to a standard NFW dark matter halo (blue dashed line). The radial variance of the fluctuations, given by Eq.~\ref{eq:fdm_var} (grey dot dashed line) is shown in arbitrary units.
         }
         \label{fig:soliton_profile}
\end{figure}
Solitons are stationary, spherically symmetric ground-state solutions of the Schrödinger–Poisson equation. They appear as a central flat density, as opposed to the cuspy inner density of NFW halos. They show a time-variable centre offset (~1 kpc) referred to as the soliton random walk, which is due to wave interference~\citep{Schive2020}. This random walk has a timescale of about $\sim100$ Myr~\citep{Schive2020}, too large to affect the lensing systems we observe. Solitons are typically characterized by a single parameter, $M_s$, the soliton mass. The so-called soliton–halo relation expresses the soliton mass in terms of its host halo and redshift. It was first found in cosmological simulations~\citep{Schive2014} and latter confirmed in other studies~\citep{Liao2024} that $M_s \propto m_{\psi}^{-1}(1 + z_h)^{1/2} M_h^{1/3}$, where $m_{\psi}$ is the axion mass, $M_h$ the halo mass, and $z_h$ its redshift. This relation has been confirmed in other studies; however, some results point towards different relations, such as $M_s \propto M_h^{5/9}$\citep{Mocz2017}. A large scatter in the soliton–halo relation has been reported, which may suggest the relation is not universal, and it remains an open question. For the rest of this work, we will assume a soliton–halo relation as found by~\citet{Schive2014} ($M_s \propto M_h^{1/3}$), specifically in the form reported by~\citet{Liao2024}, as their simulations cover higher halo masses that had not been explored in previous works.

The soliton mass density profile is given by
\begin{equation}\label{eq:soliton}
    \rho_{\rm sol}(r) = \frac{\rho_c}{\left(1+0.091\left(\frac{r}{r_c}\right)^2\right)^8},
\end{equation}
where
\begin{equation}
\rho_c = 0.019 \left( \frac{mc^2}{10^{-22}~\mathrm{eV}} \right)^{-2} \left( \frac{r_c}{\mathrm{kpc}} \right)^{-4}~M_\odot~\mathrm{pc}^{-3}.
\end{equation}
$r_c$ is the distance at which the density drops to half its central value.

The smooth component of the FDM model mass density profile, as shown in Fig.~\ref{fig:soliton_profile}, is then
\begin{equation}
\rho_{\rm FDM}(r) \simeq 
\begin{cases}
\rho_{\rm sol}(r) & r < r_{\rm sol} \\
\rho_{\rm NFW}(r) & r > r_{\rm sol},
\end{cases}
\end{equation}
where $r_{\rm sol}$ is the transition radius between models, that can be found numerically from the condition $\rho_{\rm sol}(r_{\rm sol}) = \rho_{\rm NFW}(r_{\rm sol})$ and is typically a few times $r_c$: $2.5\lesssim r_{\rm sol}/r_c \lesssim3.5$ \citep{Mocz2017, Chiang:2021uvt, Furlanetto:2024qsb}.

Finally, the 2D projection of the mass density profile (without quantum fluctuations)
can be easily obtained by numerically integrating along the line of sight
\begin{equation}
    \Sigma_{\rm FDM}(r)=2\int_0^{\infty}\rho_{\rm FDM}(r, z)\,dz.
\end{equation}
Now $r$ is defined as the radial distance within the lens plane, and $z$ is the tangential direction to such plane.
\subsubsection{Fuzzy Dark Matter fluctuations}
In the preceding subsections, we described the smooth component of the FDM profile, which is composed of a main NFW halo in the outskirts and a soliton profile in the central region. On top of that, FDM shows density clumps roughly separated by half the de Broglie wavelength, which depends on the axion and halo masses: 
\begin{equation}\label{eq:lambda_dB}
\lambda_{\rm dB} = 150 \left( \frac{10^{-22}~\mathrm{eV}}{m_\psi} \right) \left( \frac{M_h}{10^{12}~M_\odot} \right)^{-1/3}~\mathrm{pc}.
\end{equation}
Following~\citet{Amruth2023,Kawai2022,Dalal2021}, we assume that the integrated column density fluctuations along the line of sight can be well approximated as a Gaussian random field (GRF). This GRF can be drawn as a random realization from the power spectrum:
\begin{equation}\label{eq:pk}
    P(k)=\frac{4\pi}{3r_{\rm h}(x)}\left(\frac{\lambda_{\rm dB}}{2}\right)^3\exp\left(-\frac{\lambda_{\rm dB}^2k^2}{4}\right),
\end{equation}
where
\begin{equation}
    r_{\rm h}(x) =  \frac{\left( \int_Z dz\, \rho_{\rm h}(r) \right)^2}{\int_Z dz\, \rho_{\rm h}^2(r)},
\end{equation}
is the effective halo size, which contains the information of the density dispersion along the line of sight $Z$.
This field is modulated by a radial variance function that depends on $x$:

\begin{equation} \label{eq:fdm_var}
\sigma^2(x)\approx 
\left\{
\begin{array}{@{}l@{\quad}l}
\displaystyle
\frac{\kappa_s^2}{r_s \lambda_{\mathrm{dB}}}
\left[\frac{\pi}{x}-
\frac{1}{(x^2 - 1)^3}
\left(
\frac{6x^4 - 17x^2 + 26}{3}
\right. \right. \\[0.5ex]
\hspace{1.1cm}
\left. \left.
+ \frac{2x^6 - 7x^4 + 8x^2 - 8}{\sqrt{1 - x^2}} \, \mathrm{sech}^{-1}(x)
\right)
\right], & x < 1 \\[2ex]

\displaystyle
\frac{\kappa_s^2}{r_s \lambda_{\mathrm{dB}}}
\left[
\frac{\pi}{x}-
\frac{1}{(x^2 - 1)^3}
\left(
\frac{6x^4 - 17x^2 + 26}{3}
\right. \right. \\[0.5ex]
\hspace{1.1cm}
\left. \left.
+ \frac{2x^6 - 7x^4 + 8x^2 - 8}{\sqrt{x^2 - 1}} \, \sec^{-1}(x)
\right)
\right], & x > 1
\end{array}
\right.
\end{equation}
where the limit at $x\equiv r/r_s=1$ is $\frac{\kappa_s^2}{r_s \lambda_{\mathrm{dB}}}
(\pi-64/21)$ and $\kappa_s\equiv\rho_sr_s/\Sigma_{\rm crit}$.

The surface mass density of the FDM can then be drawn as a random realization given the $P(k)$ in Eq.~\ref{eq:pk} modulated by the radial variance given by Eq.~\ref{eq:fdm_var}.
Neither the soliton nor the GRF density fluctuations admit analytical expressions for their deflection angles and must therefore be computed numerically. 
\subsection{S\'ersic profile}
The profiles shown so far describe dark matter halos with good agreement to the data. Such halos account to $\sim90\%$ of the total mass of the galaxies, while the other 10$\%$ is constituted by baryons, whose mass we can trace from the light distribution. The effect of the baryons is a dampening in the FDM mass surface density fluctuations, as found by~\citet{Amruth2023}. For this work we consider only a not so large dampening factor of $\sim20\%$ to show the differences between CDM and FDM.

Surface brightness of elliptical galaxies, bulges, and disks of spiral galaxies are best fitted by a S\'ersic profile ($r^{1/n}$), where $n$ is a free parameter known as the S\'ersic index $n$. While not a 3D mass profile, the S\'ersic profile is a good approximation of the 2D projection of the mass distribution, assuming a homogeneous light-to-mass ratio. Particularly, this circularly symmetric mass profile reads as:
\begin{equation}\label{eq:sersic}
    \Sigma(r) = \Upsilon\, I_e \exp\left\{ -b(n) \left[ \left( \frac{r}{r_e} \right)^{1/n} - 1 \right] \right\},
\end{equation}
where $\Upsilon$ is the light-to-mass ratio of the galaxy, $I_e$ is the luminosity density at the effective radius $r_e$, and $b(n)$ is a constant defined so that the luminosity enclosed within $r_e$ equals half of the total luminosity. The S\'ersic index $n$ determines the concentration of the profile, with lower values corresponding to shallower inner slopes and steeper outer fall-offs. Typical values include $n \approx 1$ for exponential discs in spiral galaxies, and $n \approx 4$ for bulges and elliptical galaxies, corresponding to the de Vaucouleurs profile.

As we did for the NFW profile, we can define a dimensionless quantity $x = (r/r_e)^{1/n}$. As shown in Eq.~\ref{eq:phi_conv}, the convergence is half the Laplacian of the lensing potential. One can obtain $\psi$ by solving that equation, and the deflection angle is then the gradient of the lensing potential. To derive an analytical expression for the S\'ersic profile, we follow the procedure of~\citet{Cardone2004}. The deflection angle takes the form:
\begin{equation}\label{eq:sersic_alpha}
    \alpha(x) = 2\alpha_e\, x^{-n} \left[1 - \frac{\Gamma(2n, bx)}{\Gamma(2n)}\right],
\end{equation}
\begin{figure*}[!htp]
    \centering
    \includegraphics[width = .85\linewidth]{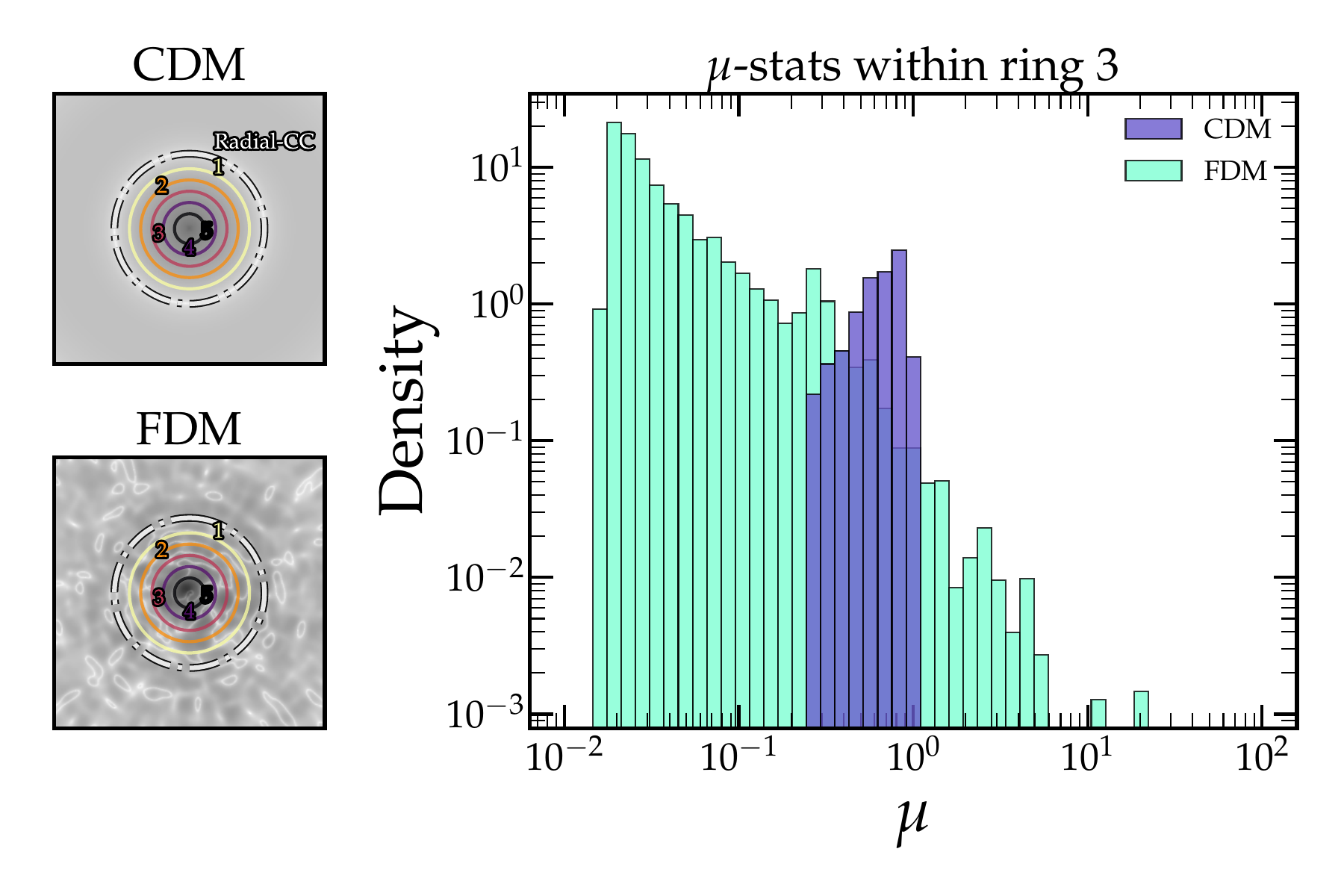}
    \caption{Depiction of this work's methodology. {\it \bf Upper left:} Zoom in into central halo region magnification map. The dot-dashed white line shows the position of the radial CC, where maximum magnification is achieved (together with the tangential CC, not shown in this plot). {\it \bf Bottom left:} Same as upper left but for the case of FDM with a soliton structure in the centre, further demagnifying the central region, and wave-like mass density fluctuations, both positive and negative. Negative fluctuations ($\delta\kappa<0$) originate islands of high magnification inside the CCs where adding more mass ($\delta\kappa>0$ both in FDM and in CDM) produces the opposite effect and increases demagnification. In both cases the coloured rings show isomagnification contours (defined in the symmetric CDM case as the CC shape and scaled down by a factor $r/r_{\mathrm{CC}}$) inside the radial-CC, where we collect the magnification statistics of the simulated pixels within each curve for later comparison. {\it \bf Right:} Example of magnification statistics. Both histograms show the magnification of the pixels within the third ring (dark-pink) of the left panels. Purple histogram shows the CDM prediction, where $\mu\lesssim1$, as expected for an NFW profile. Light-blue bins represent the FDM prediction, where both the positive and negative fluctuations in $\delta\kappa$ reduce and enhance the magnification (when compared to CDM). This translates into a high probability of both demagnified and highly magnified images. These simulations correspond to a perfectly symmetrical lens ($e=0$), a halo mass of $M_{200}=M_{h}=7\times10^{11}\,M_\odot$, and an axion mass of  $m_\psi=10^{-22}\mathrm{eV}$.}
    \label{fig:methodology1}
\end{figure*}

\vspace{-3em} 

\begin{figure*}[!!!!!!!htp]
    \centering
    \includegraphics[width = .9\linewidth]{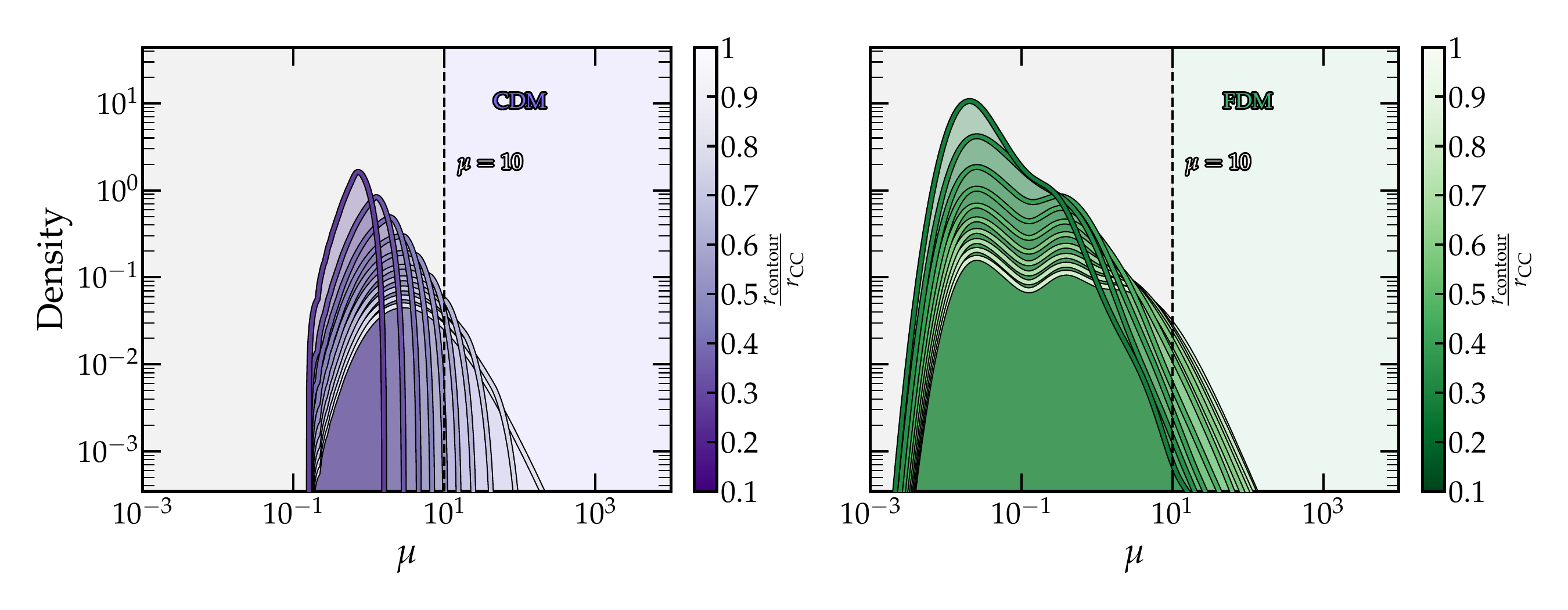}
    \caption{Probability of magnification estimated within smooth CDM isomagnification contours for the CDM profile ({\it \bf Left}) and the corresponding FDM simulations ({\it \bf Right}). Brighter colours represent larger contours, with the outermost matching the radial CC, while darker colours correspond to smaller contours closer to the centre of the halo. The dashed black vertical line marks a magnification value of 10, which we set as the threshold for comparing $p$-values.}
    \label{fig:kdes}
\end{figure*}
where $\alpha_e$ is the deflection angle at $r = r_e$, whose value is given by:
\begin{equation}
     \alpha_e =nr_e\kappa_eb^{-2n}{\rm e}^b\Gamma(2n),
\end{equation}
in arcsec units. $\kappa_e = \Upsilon\,I_e/\Sigma_{\rm crit}$, where $\Gamma(a, z)$ is the incomplete gamma function, and $\Gamma(a)$ is the actual gamma function. The parameter $b(n)$ can be found from the equation
\begin{equation}
    \Gamma(2n,b)=\Gamma(2n)/2.
\end{equation}
Tabulated values of $b(n)$ for $n \in {\{1, \ldots, 15\}}$ can be found in \citet{Mazure2002}.
\begin{figure*}[tpb]
    \centering
    \includegraphics[width = 1\linewidth]{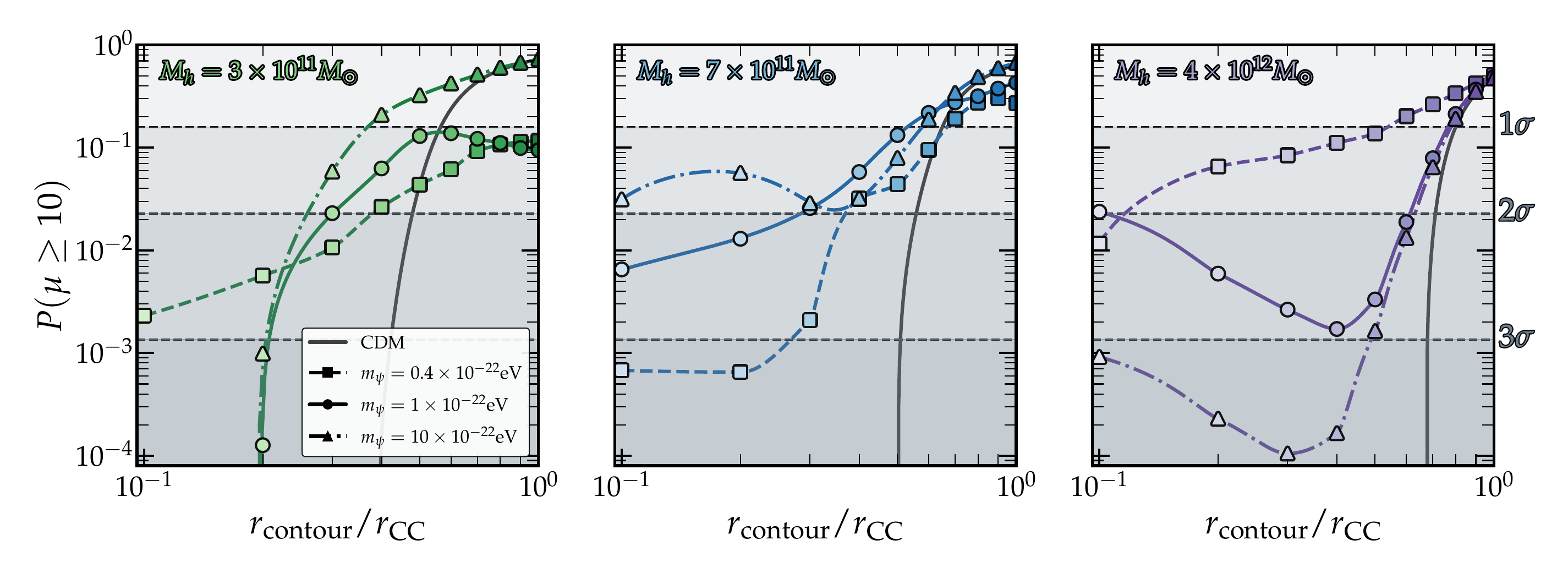}
    \caption{$p$-values for a magnification factor equal to or larger than 10 for each of the simulated lenses. The left panel shows the $p$-values for the lightest halo mass ($3\times10^{11}\,\mathrm{M_\odot}$), for all the axion masses. Middle and right panels are the equivalent $p$-values for the $7\times10^{11}\,\mathrm{M_\odot}$ and $4\times10^{12}\,\mathrm{M_\odot}$ halos, respectively. The grey lines represent the complementary $p$-values for the CDM model at each halo mass. Halo masses increase from left to right with green, blue and purple colours. The squares, circles, and triangles represent different axion masses, increasing in this particular order. Horizontal dashed lines represent the significance of finding an image with magnification equal to or larger than ten for each model. For the smooth model, it is virtually impossible to achieve such magnification at a distance to the centre of the halo  half that of the CC.}
    \label{fig:p_values_final}
\end{figure*}
\section{Methodology} \label{sec:methodology}
In this work we assume different mock lenses, varying their halo masses, ellipticities, and adding different small-scale perturbers to study the differences in radial magnification arcs between CDM and FDM for different axion masses. For each case, we compute the deflection angles, either numerically or, when possible, analytically. We then add them linearly and compute the magnification in the lens plane following Eq.~~\ref{eq:jacobian1}:
\begin{equation}
\det \tens A =
\left(1 - \frac{\partial \alpha_x}{\partial \theta_x}\right)
\left(1 - \frac{\partial \alpha_y}{\partial \theta_y}\right)
- \frac{\partial \alpha_x}{\partial \theta_y}
  \frac{\partial \alpha_y}{\partial \theta_x}.
\end{equation}
\begin{table}[tpb]
\caption{List of models.}
\label{tab:models-list}
\centering
\begin{tabular}{cccc}
\hline\hline
Model ID & $M_{h}$ [$10^{12}M_\odot$] & $m_\psi$ [$10^{-22}\mathrm{eV}$] & $\lambda_{\rm dB}$ [pc] \\
\hline
11 & $0.3$ & $0.4$ & 560 \\
12 & $0.3$ & $1$ & 224 \\
13 & $0.3$ & $10$ & 22 \\
21 & $0.7$ & $0.4$ & 422 \\
22 & $0.7$ & $1$ & 169 \\
23 & $0.7$ & $10$ & 17 \\
31 & $4$ & $0.4$ & 236 \\
32 & $4$ & $1$ & 95 \\
33 & $4$ & $10$ & 9 \\
\hline
\end{tabular}
\tablefoot{%
Model ID consists of two indices representing halo mass and axion mass, in that order, with values increasing from 1 to 3. For example, the most massive halo with the medium axion mass would be identified as 32, while the lighter halo with the heaviest axion would be indexed as 13.%
}
\end{table}
The deflection angles for the Sérsic and NFW profiles are obtained analytically (Eqs. \ref{eq:NFW_alpha} and \ref{eq:sersic_alpha}), significantly speeding up the process. The soliton and FDM fluctuations are computed numerically, according to Eqs. \ref{eq:sigmas_lin} and \ref{eq:alphas_lin}. Each pixel is treated as a point lens, and the total deflection angle is estimated via a convolution between the mass and distance kernels, using the fast Fourier transform \citep{Cooley1965}. For each case, we simulate a field of view (FOV) from 3 to 7 times the Einstein radius of the lens to fully cover the CCs. This variable FOV is motivated by the ellipticity of the lens, which stretches the CCs along one axis while compressing them along the other. We assume that all lenses are circularly symmetric to compute the deflection angle, and add ellipticity afterwards as:
\begin{equation}\label{eq:elliptical}
\begin{aligned}
F_{x_{\mathrm{elliptical}}} &= F \cdot \frac{x}{(1 - e)\, r} \\
F_{y_{\mathrm{elliptical}}} &= F \cdot \frac{y (1 - e)}{r}
\end{aligned}
\end{equation}
where $F$ is a circularly symmetric function, $x$ and $y$ are the Cartesian coordinates of the lens plane, $r \equiv \sqrt{x^2 + y^2}$ is the radial distance within the plane, and $e \equiv 1 - b/a$ is the ellipticity, with $a$ and $b$ representing the major and minor semi-axes of the ellipse.

First, for a mock lens we construct the smooth CDM lens model, combining an NFW dark matter halo with a S\'ersic profile representing the baryonic component of the galaxy. We then compute the deflection angles for each component, add them linearly, and obtain the resulting magnification map. Additionally, we assume a fiducial axion mass. For practical reasons, we select $m_{\psi}$ from three values, $m_\psi \in {0.4, 1, 10} \times 10^{-22}\,\mathrm{eV}$: the lower limit is motivated by the constraint from~\citet{Chiang2023}, as a conservative constraint on the axion mass. The middle value is chosen for historical reasons, as it is the most commonly adopted in ultralight axion FDM studies, and is also what many studies find to reproduce the observed properties of multiply-lensed images~\citep{Amruth2023,Laroche:2022pjm} or innermost kinematics of dwarf galaxies~\citep{Broadhurst2020}. The upper value lies in a regime where current constraints are still weak.

We consider three different halo masses, $M_{200}$: a fiducial mass of $7\times10^{11}\,M_\odot$ as in~\citep{Amruth2023}, a smaller halo of $3\times10^{11}\,M_\odot$, and a massive one of $4\times10^{12}\,M_\odot$. The combination of three axion and three halo masses yields a total of 9 models, as shown in Tab.~\ref{tab:models-list}. We also follow the lens parameters from~\citet{Amruth2023} for the HS~0810+2554 system, adopting $z_l = 0.89$, $z_s = 1.51$, $r_s = 50\,\mathrm{kpc}$, and $c = 9$ for $M_{200} = 7\times10^{11}\,M_\odot$, scaling them with mass as $c \propto M^{-0.1}$ and $r_{200} \propto M^{1/3}$, with $r_s = r_{200}/c$.

To test the ellipticity effect, we adopt the values $e = 0$ (a perfectly axisymmetric system), $e = 0.2$ as in~\citet{Amruth2023}, and $e = 0.4$ (an extremely elliptical configuration), for one of the simulated lenses. We first compute the analytical deflection quantities for the circular case and then apply the ellipticity transformation described in Eq.~\ref{eq:elliptical}.

For each case, once the magnification is computed, we compare the magnification statistics in the lens plane from the FDM model against the CDM model, both without subhalos (smooth case) and with small-scale perturbers. We adopt the same contour radius given by the isomagnification contours from the smooth CDM model (as shown in Fig.~\ref{fig:methodology1}) and obtain the magnification histograms of the pixels enclosed within them, as portrayed in Fig.~\ref{fig:kdes}. These isomagnification contours trace the shape of the radial critical curve in the smooth CDM model and are normalised by a factor $r/r_{\mathrm{CC}}$, where $r$ is the distance from the centre to the contour and $r_{\mathrm{CC}}$ is the distance to the critical curve. In the cases where ellipticity is introduced, $r$ and $r_{\mathrm{CC}}$ are measured with respect to the closest point on the critical curve to the centre. Once we have the histograms for the magnification inside the isomagnification contours, we can estimate the probability of having high magnification ($\mu>10$) inside the enclosed area $P(\mu \geq \mu_\varepsilon \mid \theta_{\rm model})$, i.e. the $p$-value associated with the model parameters $\theta_{\rm model}$. For an accurate depiction of the magnification statistics in the presence of FDM density fluctuations, we ensure that the pixel size provides a resolution of at least 10 pixels across the span of $\lambda_{\rm dB}$, as defined in Eq.~\ref{eq:lambda_dB}. We also perform several GRF realizations to obtain a robust estimate of the magnification statistics and to mitigate possible systematic biases arising from a single realization.
For the soliton parameters, we use the publicly available code {\tt SHR}\footnote{\url{https://github.com/calab-ntu/fdm-soliton-halo-relation}}, adopting the updated model by~\citet{Liao2024} based on the original formulation of~\citet{Schive2014}.

\section{Results} \label{sec:results}
\begin{figure}[tpb]
    \centering
    \includegraphics[width = 1\linewidth]{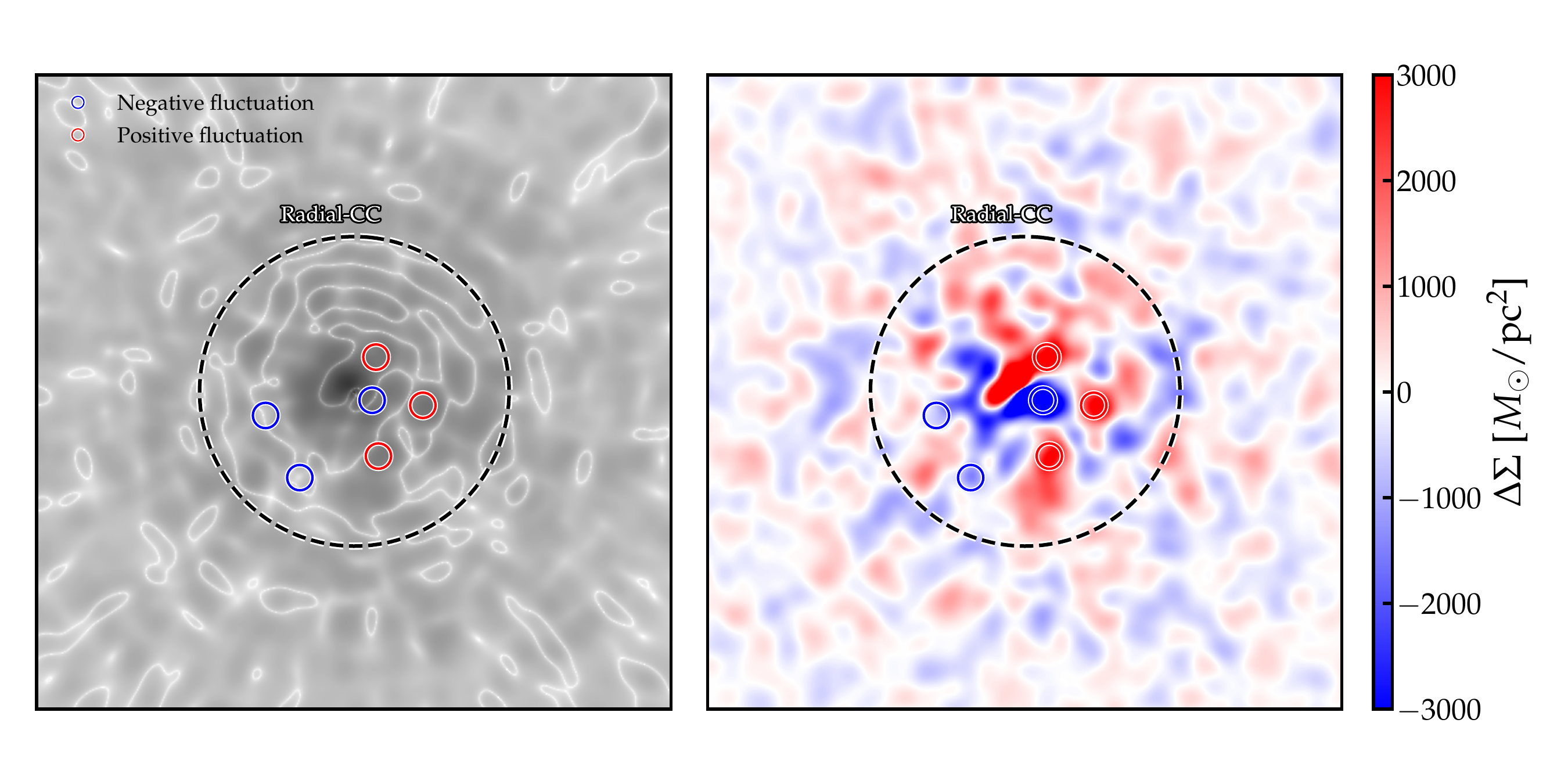}
    \caption{Magnification patterns inside the radial CC ({\it \bf left}) and mass density fluctuations ({\it \bf right}). Open blue circles mark the positions of negative mass density fluctuations, while red open circles indicate positive fluctuations. Positive fluctuations in the mass (both present in CDM and FDM) inside the radial CC lead to demagnification, whereas negative mass fluctuations (exclusive in FDM) can result in new critical regions. As we approach the centre of the halo, the negative fluctuations in FDM can compensate the increase in $\kappa$ from the underlying NFW+S\'ersic, fulfilling again the radial criticality condition, $1 - \kappa + \gamma \approx 0$.}
    \label{fig:fluct2mag}
\end{figure}

\begin{figure*}[tpb]
    \centering
    \includegraphics[width = 1\linewidth]{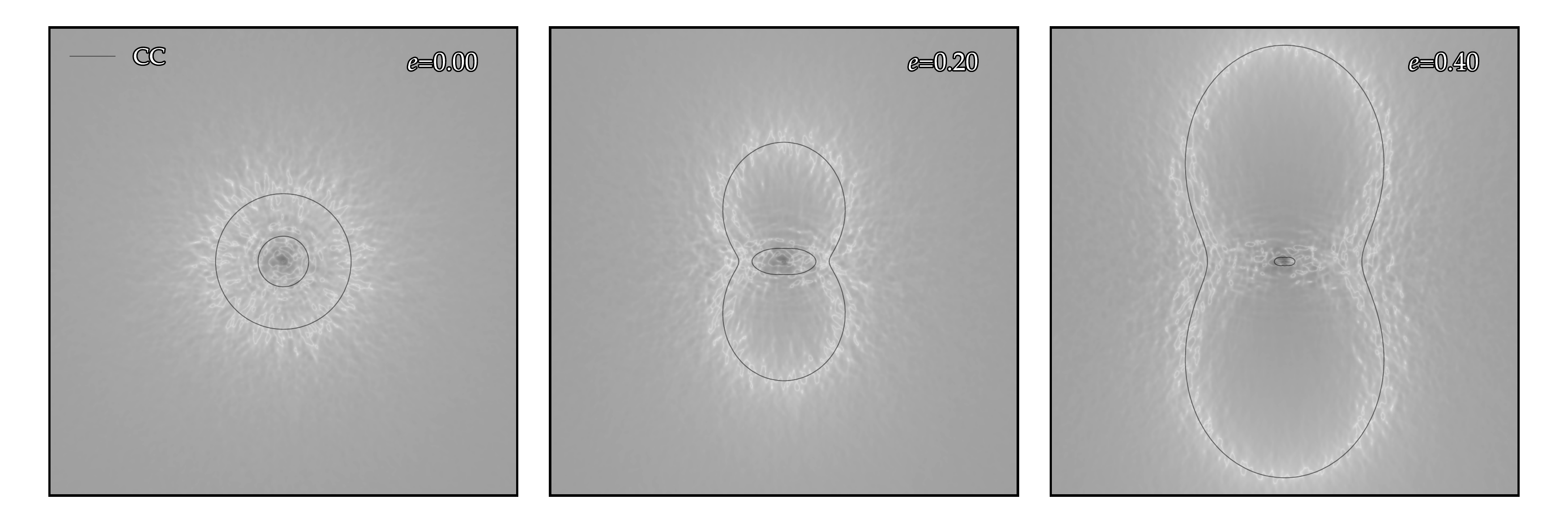}
    \caption{Ellipticity effects on magnification. From left to right: no ellipticity to highly elliptical lens. The lens parameters are those of model 22 according to table~\ref{tab:models-list}($M_{h} = 7 \times 10^{11}\,\mathrm{M}\odot$ and $m_\psi = 10^{-22}\,\mathrm{eV}$). The smooth CCs stretch along the major axis and compress along the minor axis, forming an hourglass-shaped configuration. FDM fluctuations follow the elliptical mass distribution of the lens; as a result, fluctuations located farther from the centre are attenuated, while those closer in are enhanced. Inside the radial CC, however, as the CC approaches the halo centre, the density fluctuations overlap with the soliton region and are consequently suppressed. In contrast, the interface between the tangential and radial CCs experiences enhanced magnification. The black lines correspond to the CCs of the smooth CDM model.}
    \label{fig:mu_ellip}
\end{figure*}

\begin{figure}[!htp]
    \centering
    \includegraphics[width = 1\linewidth]{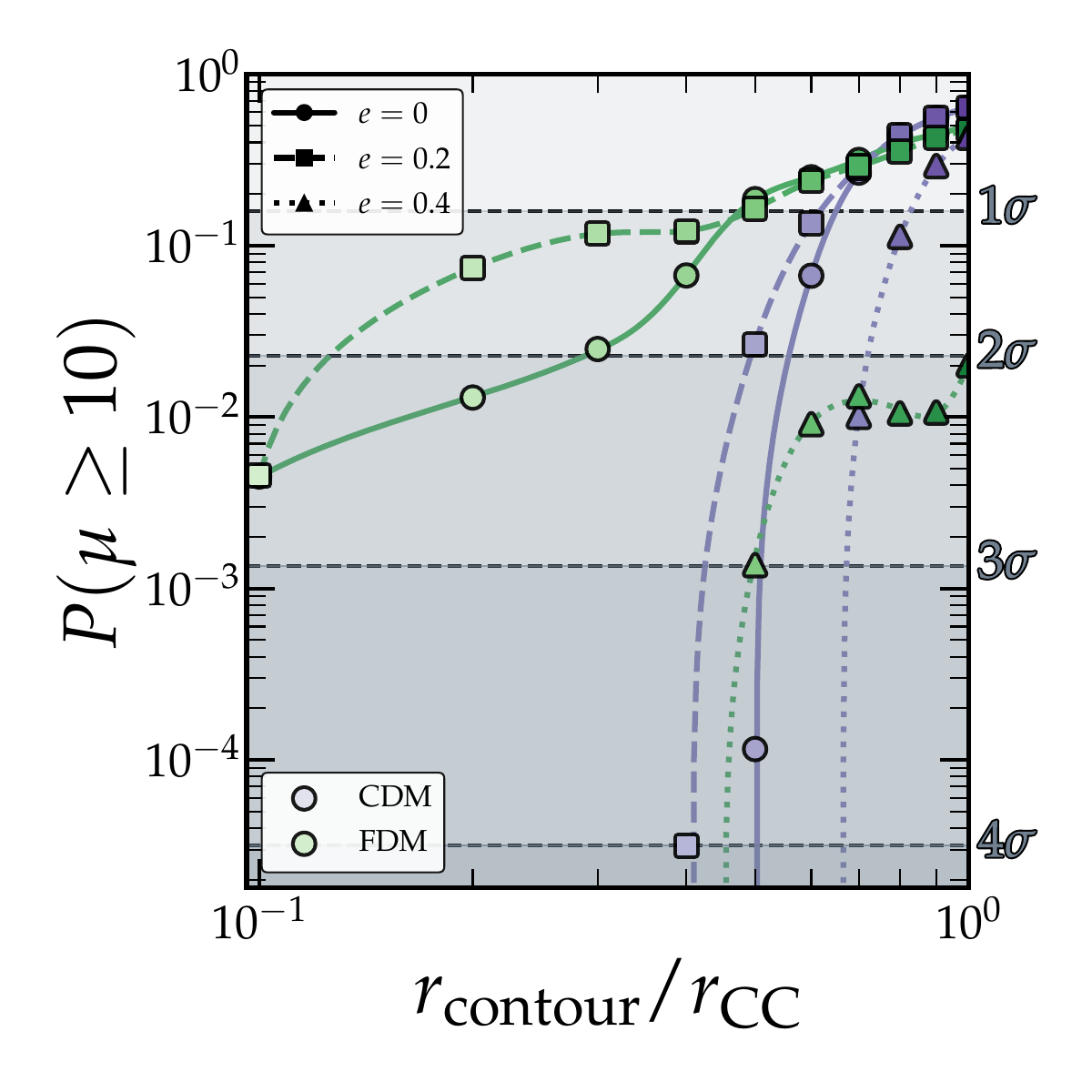}
    \caption{$p$-values for magnification values equal or larger than 10 in both the smooth CDM model (in purple), and the wave-like FDM model (in green). Each point represents the $p$-value for the pixels inside the scaled CC contours with a size given as a fraction of the radial CC size ($r/r_{\mathrm{CC}}$). Different marker symbols represent the $p$-values of the different ellipticity realizations. As in Fig.~\ref{fig:p_values_final}, the $p$-value is the probability within the given model of attaining a magnification equal or larger than 10, and the horizontal dashed lines show the significance contours. The halo and axion masses correspond to model 22 in Table\ref{tab:models-list}, with $M_{h} = 7 \times 10^{11}\,\mathrm{M}\odot$ and $m_\psi = 10^{-22}\,\mathrm{eV}$.}
    \label{fig:pvalues_ellip}
\end{figure}

In this section, we present the main results from the analysis of magnification distributions within radial CCs in FDM-simulated lenses (see Fig.~\ref{fig:p_values_final}), and compare them with their CDM counterparts. We also examine the impact of ellipticity in both scenarios, and finally explore the effects of adding NFW-subhalos to the CDM model compared to the FDM case. For reference, the magnification maps corresponding to the lens models listed in Tab.~\ref{tab:models-list} are shown in Appendix~\ref{app:lenses}.

The key distinguishing feature of FDM (and $\psi$DM in general) compared to other dark matter models is that the destructive interferences give rise to negative density fluctuations with respect to the NFW baseline, which in turn generate high-magnification regions inside the CCs (see Fig.~\ref{fig:fluct2mag}). Within these curves, the critical condition is fulfilled, $1 - \kappa - \gamma \approx 0$ for tangential CCs and $1 - \kappa + \gamma\approx0$ for radial CCs. Adding mass in the form of small-scale perturbers can modify the local magnification distribution, but, on average, it does not significantly affect the global statistics, except for small background sources that are near to the perturbers, where the local PDF can change substantially. In contrast, the wave-like fluctuations inherent to FDM can produce extended high-magnification regions (as long as the associated de Broglie wavelength is sufficiently large relative to the lensing scale). Said regions are intrinsically different from those produced by CDM subhalos or other small-scale structures. In the regime where the fluctuation scale is much smaller than scale of the lens, given by its Einstein radius, such as in galaxy clusters, FDM behaves effectively as a population of millilenses~\citep{Diego2024,Perera2025}.

The two main parameters that govern the FDM fluctuations and thus the changes in magnification statistics are the halo mass, $M_h$, and the axion mass, $m_\psi$. The de Broglie wavelength scales with both parameters as shown in Eq.~\ref{eq:lambda_dB}, where $\lambda_{\rm dB}$ increases as $\lambda_{\rm dB} \propto m_\psi^{-1}$ and $\lambda_{\rm dB} \propto M_h^{-1/3}$. This sets the axion mass as the most important factor in the growth of the density fluctuation scale. The size of the CCs or, in other words, the lensing area of influence for strong lensing effects, increases with the halo mass roughly as $M$. 


Models 11 and 21 feature a $\lambda_{\rm dB}$ much larger than the typical lens scale. In such cases, the resulting perturbations in the magnification patterns are excessively strong and would likely have already been detected through image position anomalies. Moreover, no Einstein rings would form in these scenarios, and the deviations far exceed existing tensions. Model 12 behaves similarly, although the effects are slightly milder than in models 11 and 21.

Models 13, 22, and 31 show a large effect without completely disrupting the CCs as in models 11, 21, and 12. These models present a central demagnification caused by the soliton, corresponding to larger halo and axion masses. All these models show the largest $p$-values according to Fig.~\ref{fig:p_values_final}.

Models 23 and 32 are similar to each other and show smaller perturbations confined to regions very close to the CCs, resembling the millilensing or microlensing regime. Their $p$-values also indicate an increased probability of high-magnification regions ($\mu \geq 10$) near the centre, due to the presence of a secondary CC created by the soliton, at the cost of enhanced demagnification further in.

Finally, model 33 illustrates the regime in which FDM approaches standard CDM within our resolution limits. At this resolution, the only distinction between FDM and CDM is the presence of the solitonic core, which increases the probability of high magnification near the halo centre. At much higher resolution, the small-scale density fluctuations would still be visible, and their effect would resemble that of microlenses.

\subsection{Ellipticity} \label{sec:ellipticity}

\begin{figure}[tpb]
    \centering
    \includegraphics[width = 1\linewidth]{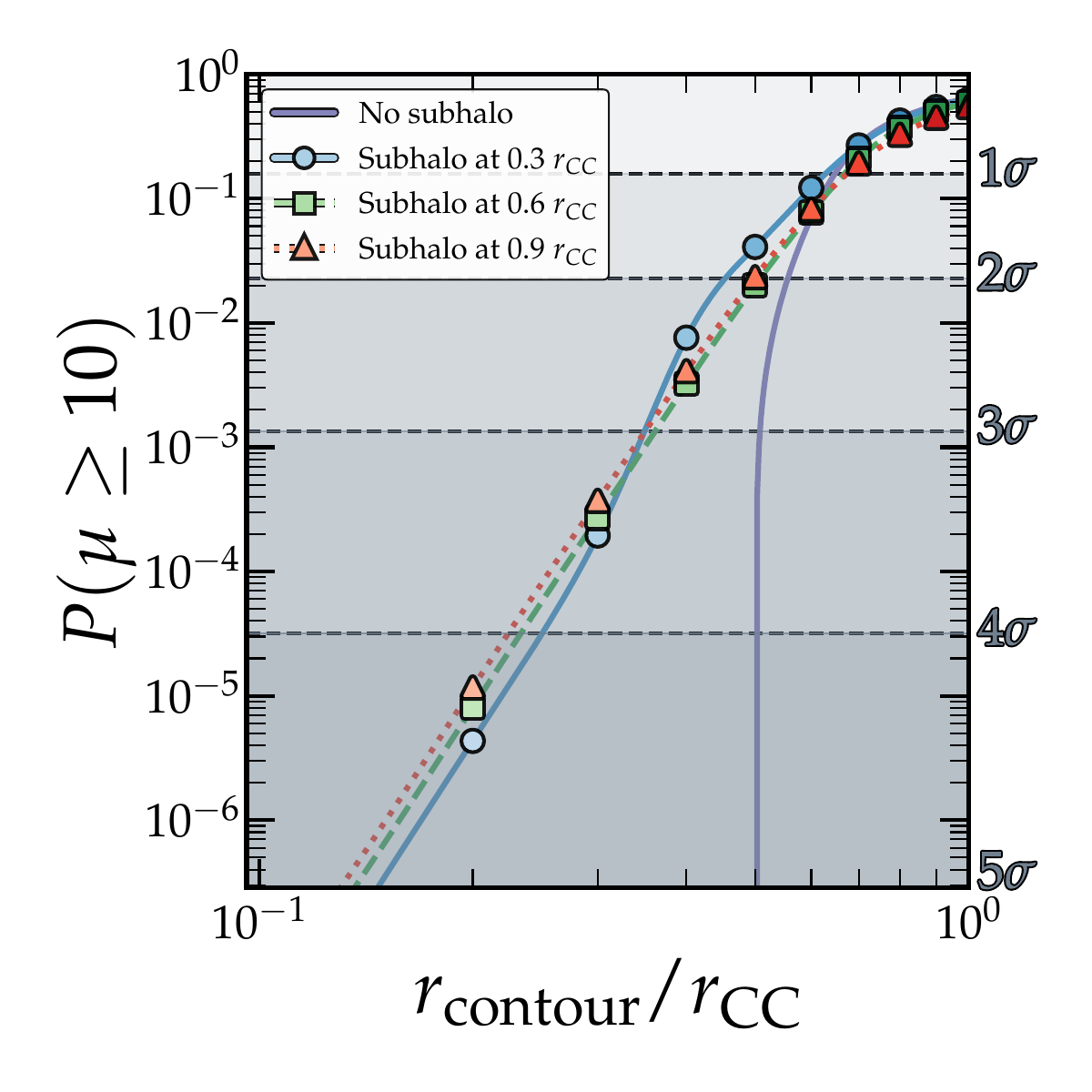}
    \caption{$p$-values for magnification values equal or larger than 10 in the smooth CDM model (in purple) and adding a subhalo of $10^7M_\odot$ at different positions with respect to the centre of the main halo. Same as Figs.~\ref{fig:p_values_final} and~\ref{fig:pvalues_ellip}, the $p$-value is the probability of obtaining an image with a magnification equal to or larger than 10 given the models, and the significance values are marked as the horizontal dashed lines. The halo and axion masses correspond to model 22 in Table \ref{tab:models-list}, with $M_{h} = 7 \times 10^{11}\,\mathrm{M}\odot$ and $m_\psi = 10^{-22}\,\mathrm{eV}$.}
    \label{fig:pvalues_fix_mass}
\end{figure}


\begin{figure}[tpb]
    \centering
    \includegraphics[width = 1\linewidth]{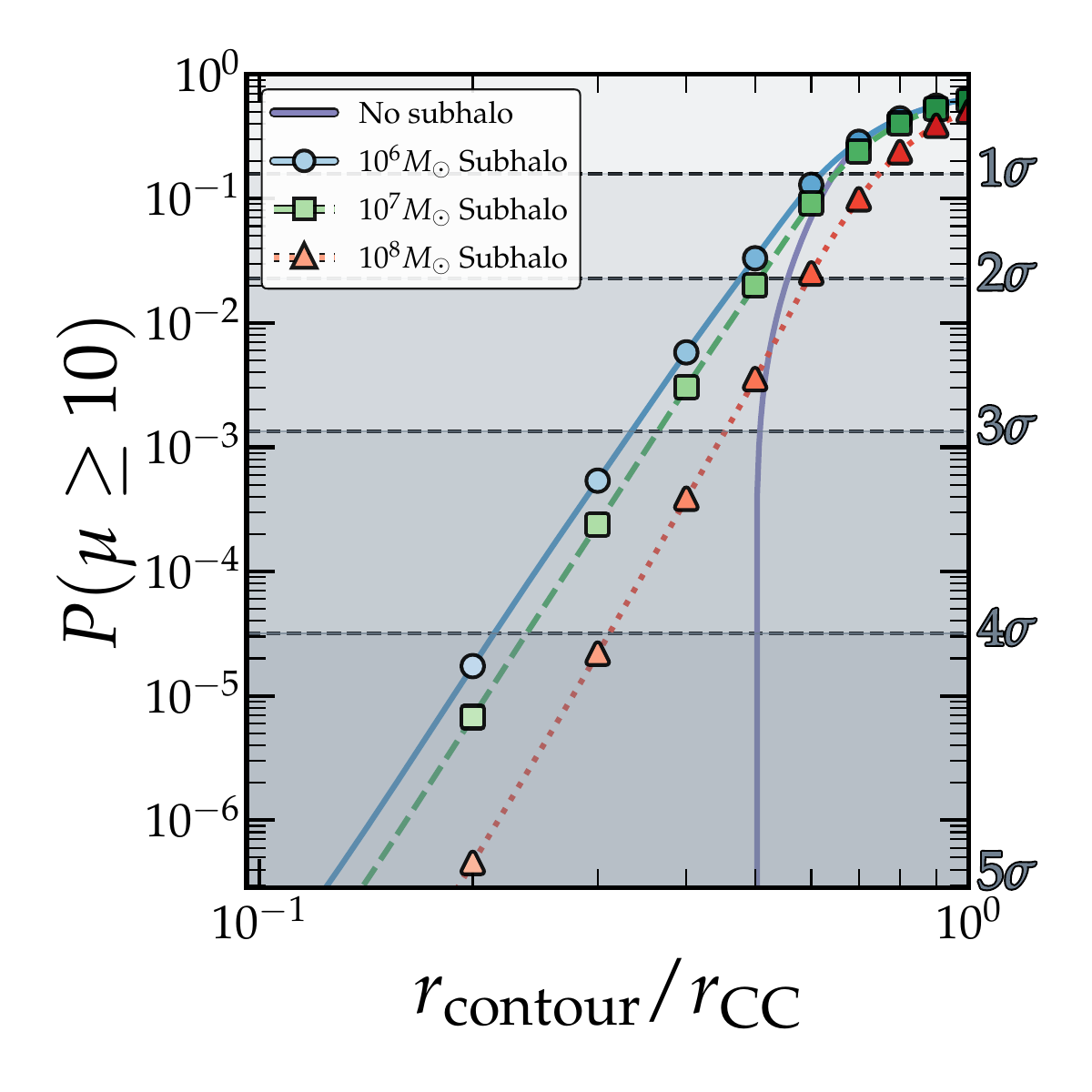}
    \caption{Same as Fig. \ref{fig:pvalues_fix_mass}, showing $p$-values for magnifications equal to or greater than 10 in the smooth CDM model and after adding subhalos. In this case, the subhalo mass varies between realisations, as colour-coded in the legend, while its position is fixed at half the radial critical curve radius from the centre of the main halo. The halo and axion masses correspond to model 22 in Table \ref{tab:models-list}, with $M_{h} = 7 \times 10^{11}\,\mathrm{M}\odot$ and $m_\psi = 10^{-22}\,\mathrm{eV}$.}
    \label{fig:pvalues_fix_position}
\end{figure}
In this work, we have only considered axisymmetric lenses up to this point. However, real lenses typically exhibit some degree of ellipticity. This ellipticity ranges from 0, corresponding to perfectly circular profiles, up to about $e = 0.5$, which is considered extremely large, as higher values are rare in nature.

For the ellipticity, we adopt three nominal values: 0, 0.2, and 0.4, and redo the same analysis in each case, studying the magnification distributions within isomagnification contours inside the radial-CC. For simplicity, we restrict ourselves to a single axion and halo mass, corresponding to those of model 22. Such magnification maps are shown in Fig.~\ref{fig:mu_ellip}. What we observe is a decrease in the area enclosed by the radial-CC, which moves progressively closer to the centre of the halo. Initially, the shape stretches along the direction of the minor semi-axis, but at higher ellipticities it also shrinks along that direction. As seen in Fig.~\ref{fig:pvalues_ellip}, for $e = 0.2$, the $p$-values are actually higher than in the circular case. At the largest ellipticity, however, the area of high magnification becomes much smaller. Images inside the smaller radial-CC are more likely to be demagnified, thus the $p$-values are smaller than the two other cases. Nevertheless, there are some regions of high magnification between CCs along the minor semi-axis, although we do not cover those in this work.

\subsection{CDM subhalos} \label{sec:subhalos}


The main objective of this paper is to show that the enhanced magnification statistics observed in the central regions of FDM halos (containing both negative and positive mass fluctuations with respect the underlying NFW) are difficult to reproduce within classical $\Lambda$CDM models (where mass offsets with respect to the NFW baseline can be only positive), and appear to be a unique feature of the negative fluctuations in FDM (or potentially other wave-like dark matter models not considered here). To explore this, we focused again on halo model 2 ($M_h=7\times10^{11}\,\mathrm{M_\odot}$) and introduce subhalos at various positions and with different masses inside the region enclosed by the radial CC.

We consider three subhalo masses: $10^6\,\mathrm{M_\odot}$, comparable to the most massive globular clusters; $10^7\,\mathrm{M_\odot}$; and $10^8\,\mathrm{M_\odot}$, representative of low-mass dwarf galaxies. Different subhalo positions are explored to account for the enhanced lensing effect when a lens lies near the CC. In such cases, a point-mass lens of mass $M$ located close to a CC with magnification $\mu$ behaves as if it had an effective mass of $M_{\mathrm{eff}} = \mu \times M$. We model subhalos with axisymmetric NFW profiles, adopting concentration parameters scaling as $M^{-0.1}$~\citep{Dutton2014}, and scale radii scaling as $M^{1/3}$. We adopt reference values of $c = 45.7$ and $r_s = 0.18\,\mathrm{kpc}$ for a $10^6\,\mathrm{M_\odot}$ subhalo, and scale these quantities accordingly for the other masses.

When including subhalos, we find a slight increase in the magnification $p$-values (for $\mu \geq 10$) compared to the smooth CDM case, as shown in Figs.~\ref{fig:pvalues_fix_mass} and~\ref{fig:pvalues_fix_position}. However, this enhancement is smaller than the one produced by FDM (see Fig.~\ref{fig:p_values_final}). Interestingly, when the subhalo is too massive (10$^8$ M$_\odot$), it can even reduce the $p$-values with respect to when considering a large radius for the contour of the studied region, compared to the base model without subhalos, as illustrated in Fig.~\ref{fig:pvalues_fix_position}. This can be understood as an increase in the curvature of the deflection field of the otherwise smooth model. Since magnification is inversely proportional to the curvature of the deflection field, adding more (positive) substructure reduces the magnification. Alternatively, this can be understood as the term $\mu_r^{-1}=1-\kappa+\gamma$ departing even more from $\approx 0$ (and becoming more negative) when  $\kappa$ is increased inside the radial CC. 

The inclusion of subhalos induces two distinct effects: (i) they can form small secondary CCs at their positions (higher $\mu$), but locally reducing magnification near them. However, the $p$-values in these regions remain low, and highly magnified central images remain statistically rare; (ii) if located near the radial CC and sufficiently massive, a subhalo can distort and pull the main CC inward. Even then, the probability of achieving high magnifications stays below that expected from FDM.

When subhalos are placed closer to the halo centre, their impact on the global magnification statistics within the radial CC region diminishes, as seen in Fig.~\ref{fig:pvalues_fix_mass}. In the limit where a subhalo lies exactly at the centre, its behaviour becomes nearly indistinguishable from the smooth CDM case. This effect is partially explained by the magnification of the macro model (i.e. the galaxy halo) at the subhalo’s location. Subhalos near the radial CC, where $\mu > 1$, thus behave as if they were more massive, effectively shifting the radial CC inward and increasing the $p$-values in the small contours near the centre. Conversely, subhalos located closer to the centre, where $\mu < 1$, behave as if they had lower effective masses. In these cases, any increase in $p$-values is primarily driven by the formation of secondary CCs, similar to those produced by a solitonic core, but these secondary CCs can also demagnify the region inside them. Since the central macro magnification is lowest, the enhancement in $p$-values for centrally placed subhalos is minimal.

When the subhalo position is fixed (see Fig.~\ref{fig:pvalues_fix_position}), the behaviour simplifies: more massive subhalos, being stronger overdensities, more effectively demagnify their central region and shrink the higher magnification area. Subhalos closer to the radial CC remain more efficient at perturbing and shifting the CC inward.

While subhalos alone cannot reproduce the magnification statistics of FDM, they may still contribute to positional anomalies, similarly to the perturbations introduced by small satellite galaxies or low-mass cluster members.

\section{Discussion} \label{sec:discussion}
The results presented in this work highlight a key observational distinction between standard CDM and FDM in the context of strong gravitational lensing: FDM gives rise to a higher probability of magnification within the radial CC as a consequence of interference-driven density patterns. In particular, the negative mass fluctuations in FDM can compensate the increase in $\kappa$ inside the radial CC, resulting in new critical regions around the negative fluctuations. Meanwhile, standard CDM models, even when including subhalos, fail to reproduce similar levels of central magnification. This behaviour, unique to FDM scenarios, offers a promising avenue for constraining the nature of dark matter through precise lensing measurements.

In particular, we find that the presence of extended high-magnification regions near the centre of lenses, associated with soliton-induced secondary CCs and fluctuations on de Broglie scales, can hardly be mimicked by conventional substructure(s) in CDM. This result suggests that the magnification statistics within radial-CCs, over a large span of well-modelled lenses, can provide a complementary probe of FDM. With the next generation of surveys, such as the ongoing {\it EUCLID} mission, the upcoming {\it Rubin-LSST}, and {\it Roman}, thousands of galaxy-scale lensing systems are expected to be discovered. These systems can then be followed up with specialized instruments that are better suited for strong lensing and/or spectroscopic analysis.

The magnification statistics presented here characterise regions of high magnification, but do not capture the spatial distribution of magnifications. Extended sources would require a follow-up analysis that would consider convolved magnification maps with specific source shapes. Nonetheless, point-like sources are actually better suited to test this statistical framework. Given the resolution of our simulation, we argue that bright, small sources such as quasars (QSOs) are well described by our results and represent an ideal observational test of the predictions presented here. Subparsec compact but luminous star clusters, which are larger sources, are even more suitable, as they are less affected by microlensing than QSOs.

While some axion mass ranges close to $10^{-22}\,\mathrm{eV}$ have been excluded by complementary constraints, some would argue that these limits remain weak due to modelling assumptions or noisy data. Here, we present a complementary lensing-based effect to constrain such ranges. 

The smallest axion masses ($10^{-23}\,\mathrm{eV}$) are unlikely, as the large distortions they produce in the critical curves should already have been observed. The intermediate range, around $10^{-22}$–$10^{-21}\,\mathrm{eV}$, yields some plausible effects. However, we argue that the largest axion mass considered may be the most compelling case. While smaller axion masses produce deviations from the smooth CDM model that are too strong to have gone unnoticed, this higher mass yields lensing effects that are still significant when applied to lower-mass halos, but remain consistent with current observational constraints in more massive halos, where the behaviour closely resembles that of CDM. Interestingly, these smaller halos are harder to detect observationally, which suggests that this axion mass still offers a viable and testable window for future studies.

There remain some unmodelled effects, such as the presence of a central supermassive black hole. We argue that such a component would act as a point-like lens, slightly shifting the smooth CC outwards and introducing a small inner CC, while leaving the main results of this work effectively unchanged. However, its impact on the solitonic structure is less clear and lies beyond the scope of this paper, so a detailed analysis is deferred to future work. It is also important to note that the true global density profiles of real galaxies remain unknown. Current models represent our best approximations based on available data and theoretical expectations. Any revision to our understanding of the smooth mass distribution would directly affect the resulting lensing predictions and constraints.

A source of uncertainty arises in the detection of central bright images, due to the difficulty in disentangling them from the emission of the lens galaxy itself. In this regard, spectroscopic follow-up observations would be valuable to distinguish between the two, as complementary images of the lensed source provide an expected spectrum that can be used to separate it from the lens galaxy. The application of this methodology to observational data, along with the associated challenges, will be addressed in future studies. With the advent of new facilities that will carry out large deep surveys, we anticipate a wealth of data, in particular for central images of lensed QSOs, that will enable a detailed statistical study of CDM vs FDM.

\section{Conclusion} \label{sec:conclusion}
We have explored the impact of wave-like dark matter fluctuations on gravitational lensing magnification patterns, with a focus on radial critical curves (CCs). Our analysis reveals observational signatures that distinguish FDM from classical CDM, particularly in compact lensing configurations. The main conclusions of this work are:
\begin{itemize}
\item We have shown that wave-like density fluctuations inherent to FDM can enhance magnification near radial CCs in a way that is difficult to reproduce with CDM, even when including substructure. In particular, the negative fluctuations in FDM (with respect to the NFW profile), which are not present in standard CDM, can produce islands of high magnification in the region interior to the radial critical curve. Such islands cannot be reproduced by the classical CDM models. 

\item Our results indicate that axion masses in the range $10^{-22}$–$10^{-21}\,\mathrm{eV}$, especially in low-mass halos, produce effects distinguishable from CDM, and merit further investigation.

\item Statistical differences in magnification, particularly for point-like sources such as QSOs or compact star clusters near the centres of lenses, offer a promising observational signature for constraining ultralight axion dark matter. This work motivates targeted searches for these compact, highly magnified sources, within radial arcs as potential probes of the FDM parameter space.

\item While enhanced central images are more likely in the FDM scenario, they remain plausible, though less probable, in standard CDM models with subhalos. To place robust constraints on the axion mass, a statistically significant sample of lenses with anomalous radial images is required. Each system would need a tailored lens model, as the uncertainties associated with the observational data are highly non-linear. A joint likelihood analysis combining multiple systems would strengthen the constraints. Upcoming wide-field surveys, such as {\it Euclid}, {\it Rubin-LSST}, or {\it Roman}, are expected to detect numerous lensed systems, providing a promising dataset to pursue this approach. The number of systems with anomalous radial images required to place a robust constraint on the axion mass within the FDM framework is beyond the scope of this paper, as each system would require a tailored lens model and propagate the highly non-linear uncertainties inherent in the observational data.

\end{itemize}

\section*{Acknowledgements}

We sincerely thank Hsi-Yu Schive and Pin-Yu Liao for insightful discussions regarding the soliton–halo relation and galaxy parameters.

JMP acknowledges financial support from the Formaci\'on de Personal Investigador (FPI) programme, ref. PRE2020-096261, associated with the Spanish Agencia Estatal de Investigaci\'on project MDM-2017-0765-20-2.
J.M.D. acknowledges the support of project PID2022-138896NB-C51 (MCIU/AEI/MINECO/FEDER, UE) Ministerio de Ciencia, Investigaci\'on y Universidades.
SKL and JL acknowledge the General Research Fund under grant RGC/GRF 17312122, which is issued by the Research Grants Council of Hong Kong S.A.R.
PM and TJB are supported by the Spanish Grant PID2023-149016NB-I00 (MINECO/AEI/FEDER,UE), as well as the Basque government Grant No. IT1628-22. PM also acknowledges financial support from fellowship PIF22/177 (UPV/EHU).
We acknowledge Santander Supercomputacion support group at the University of Cantabria who provided access to the supercomputer Altamira Supercomputer at the Institute of Physics of Cantabria (IFCA-CSIC), member of the Spanish Supercomputing Network, for performing simulations.

We made use of the following software: Python, NumPy, SciPy, Matplotlib, Astropy, and Powerbox.

\bibliography{bibtex}{}
\bibliographystyle{aasjournal}

\appendix
\onecolumn
\section{Simulated lenses}\label{app:lenses}
\begin{figure*}[h!]
    \centering
    \includegraphics[width=1\textwidth]{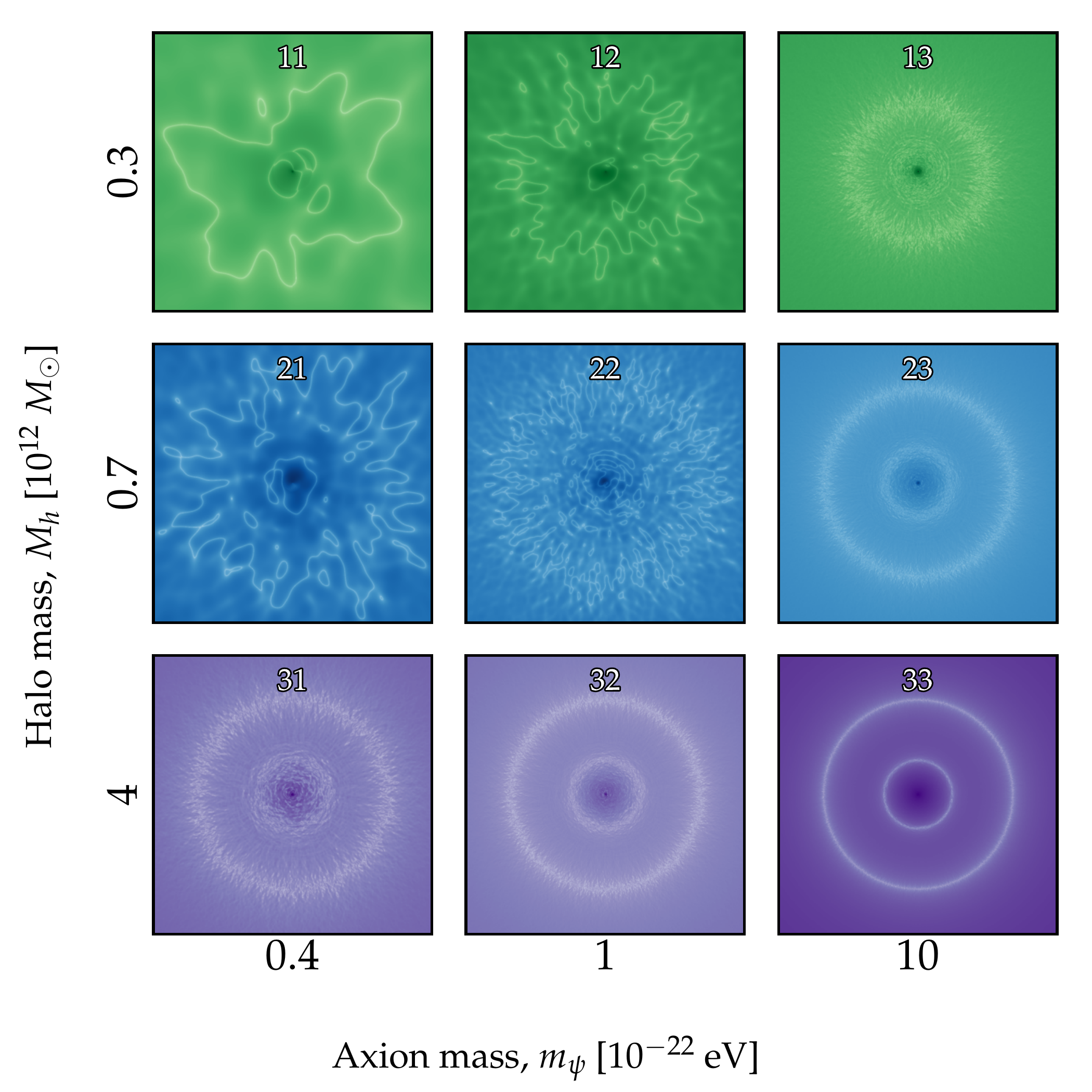}
    \caption{Magnification maps of simulated lenses according to Tab.~\ref{tab:models-list}. Axion mass increases from left to right, and halo mass increases from top to bottom.}
    \label{fig:lenses}
\end{figure*}
\clearpage
\section{Image position anomalies}\label{app:image_anomalies}
\begin{figure}[h!!!!!!!!!!!!!!!!!!!!!]
    \centering
    \includegraphics[width=1\textwidth]{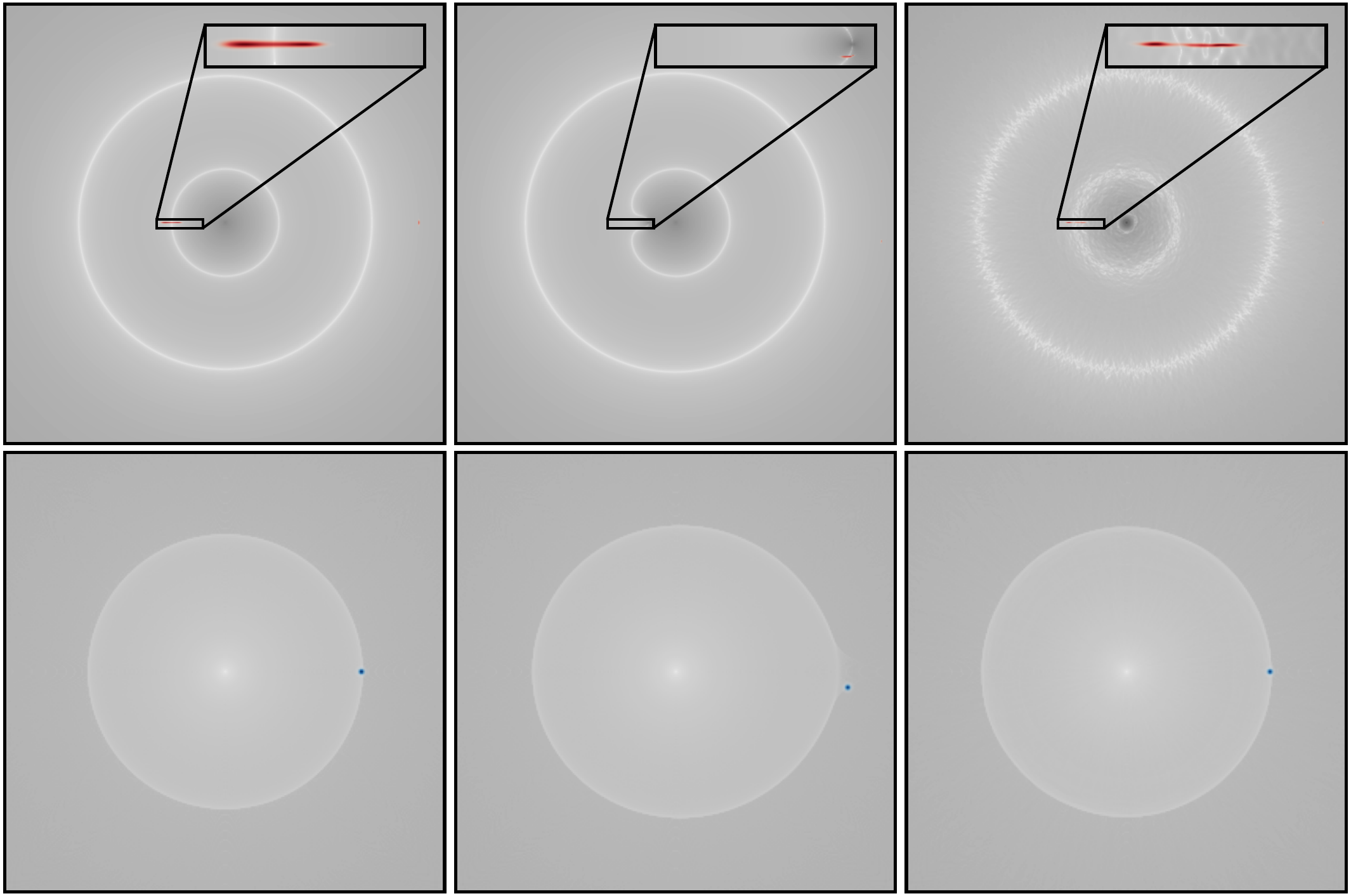}
    \caption{Radial image configurations for three lens models: a smooth particle-like CDM model (left panel), the same model with a $10^7\,\mathrm{M}_\odot$ subhalo placed halfway between the centre and the radial CC (central panel), and an FDM model corresponding to an axion mass of $5\times10^{-22}$ eV (right panel). The top row shows the CCs, while the bottom row shows the corresponding caustics. The blue dots in the bottom panels represent a Gaussian source with a width of 0.8 pixels, placed at the radial caustic to produce bright radial images on the CC. The resulting arcs or images appear in red in the top panels. In the smooth CDM case, the arcs are symmetric and located on top of the radial CC. When a subhalo is added, a similar configuration arises, though the image appears closer to the centre and with lower magnification. In the FDM case, the arcs are asymmetric and exhibit magnification fluctuations on the scale of the axion’s de Broglie wavelength. The two merged radial images seen in the CDM case are now split into distinct components by the fluctuations. Although the caustics are broader in this case, we can confidently say that for an FDM lens, if the source size is comparable to the de Broglie scale, significant image position asymmetries and brighter images closer to the centre can arise compared to the particle-CDM model.}
    \label{fig:image_anomalies}
\end{figure}

\end{document}